\documentstyle[prl,aps,preprint,tighten,floats,aps,epsf,psfig]{revtex}

\def\be{\begin{equation}}
\def\ee{\end{equation}}
\def\bear{\begin{eqnarray}}
\def\eear{\end{eqnarray}}
\def\beqn{\begin{eqnarray}}
\def\eeqn{\end{eqnarray}}

\def\e{{\rm e}}

\def\MP{M_{Pl}}

\def\beq{\begin{equation} }
\def\eeq{\end{equation} }

\def\ben{\begin{eqnarray} }
\def\een{\end{eqnarray} }

\def\mod#1{{\rm (mod~2)} }

\def\Tr{{\rm Tr}}

\def\Tr{{\rm Tr}\,}

\begin{document}
\draft
\preprint{\vbox{\baselineskip=12pt
\rightline{CERN-TH/97-338}
\vskip0.2truecm
\rightline{UPR-0779-T}
\vskip0.2truecm
\rightline{IEM-FT-166/97}
\vskip0.2truecm
\rightline{hep-th/9711178}}}

\title{Classification of Flat Directions  in Perturbative  Heterotic
 Superstring Vacua with Anomalous $U(1)$} 
\author{Gerald Cleaver${}^{\dagger}$, Mirjam Cveti\v c${}^{\dagger}$, 
Jose R. Espinosa${}^*$${}^{\dagger}$, Lisa Everett${}^{\dagger}$, and
Paul Langacker${}^{\dagger}$}
\address{${}^{\dagger}$Department of Physics and Astronomy \\ 
          University of Pennsylvania, Philadelphia PA 19104-6396, USA \\
${}^*$CERN, TH Division\\
CH-1211 Geneva 23, Switzerland}
\maketitle
\begin{abstract}
We  develop techniques to  classify
$D$- and $F$-flat directions  for $N=1$ supersymmetric  string vacua of the
perturbative heterotic string theory, which  possess 
 an anomalous $U(1)$ gauge group at the tree level. Genus-one corrections  
generate a
 Fayet-Iliopoulos term for the $D$-term of $U(1)_A$, 
 which is canceled by 
 non-zero  vacuum 
  expectation values (VEVs) of certain 
massless  multiplets in such a way  that the anomalous $U(1)$ is broken,
 while
  maintaining the  $D$- and  $F$-flatness of  
  the  effective field theory. A systematic analysis of flat
  directions is  given for non-zero VEVs of non-Abelian singlets,  and
 the techniques are illustrated for a specific model.  The approach  
sets the
stage to  classify 
the $D$- and $F$-flat  directions for a large class 
of perturbative   string vacua. This classification is
 a prerequisite to address systematically the phenomenological consequences of
 these models.
\end{abstract}
\vskip2cm
\leftline{} 
\leftline{CERN-TH/97-338}
\leftline{November 1997}  
\newpage
\section{Introduction}
As a part of a broader program to address phenomenological implications 
 of  string models one is faced with a myriad of challenges, among others the
 degeneracy of string vacua,  origin of supersymmetry breaking, and for some
 string vacua  calculability of the effective Lagrangian.

Nevertheless, one can take a more modest approach  to consider
 a class of specific  
superstring vacua  which at  the string scale
 ($M_{string}$) possess  $N=1$
 supersymmetry,   the standard  model (SM) gauge group 
  $SU(2)_L\times U(1)_Y\times SU(3)_C$,  
as a part of the
full gauge structure, and a particle content that includes 
 three ordinary families 
 and  at least two SM
Higgs doublets; i.e., such superstring vacua possess 
at least the ingredients of the  minimal supersymmetric standard model (MSSM).
These models 
may provide  candidate vacua with 
specific phenomenological predictions, 
consistent with the MSSM. 

Within the {\it perturbative heterotic superstring}, classes of such
quasi-realistic superstring models were constructed from orbifold
constructions \cite{DHVW} with Wilson lines \cite{Wilson,orbifolds},
Calabi-Yau
compactifications \cite{calabiyau}, free-fermionic constructions
\cite{ff,fermionic,CHL}, covariant lattices \cite{covariantlattice} and
Gepner-Kazama-Suzuki models \cite{GKZ}. 
 A set of these  models constitutes a starting point to address  their detailed
phenomenological analysis.
In general they possess  the following features:
\begin{itemize}
\item 
{\bf Gauge Group Structure.}
Along with the SM gauge group there is  a  non-Abelian hidden 
sector group and a  number of  additional $U(1)$'s, one of them 
generically {\it anomalous}. The hidden sector non-Abelian gauge group  may 
play a role in
dynamical supersymmetry breaking.

The $U_Y$ is  determined as a linear combination
of the $U(1)$ factors (or possibly broken  hidden sector non-Abelian factors), 
subject to the constraint that the massless spectrum
contains the MSSM  particles (three ordinary families and two Higgs doublets),
and  the pairing of the  exotic  $SU(3)_C$ triplets, $SU(2)_L$ doublets and 
exotic singlets of the observable sector~\footnote{A weaker requirement is that
the exotic particles are paired with respect to $SU(3)_C$ and the electric
charge quantum number, so that there will be no {\it exactly} massless colored 
or
electrically charged fermions in the theory. For this weaker requirement there 
can be
exotic states that are chiral under $SU(2)_L$ and/or $U(1)_Y$, but can acquire
masses at the electroweak breaking scale.}.  These constraints  significantly 
restrict the
allowed values of $U(1)_Y$, often imposing a
unique choice of  $U(1)_Y$.

\item 
{\bf Particle Spectrum.}
The mass spectrum is calculable.  In addition to the  three
ordinary families and two SM Higgs doublets, there  are generically a
 large number of 
 exotic  massless states. In particular,
 there are usually a large number of additional massless matter multiplets,
which  transform non-trivially under the $U(1)$'s  and/or the  standard 
model  symmetry, as well as hidden sector states that transform non-trivially
under  the $U(1)$'s and the non-Abelian  hidden sector gauge group.
 In some cases these models  also possess massless
   states that   are non-Abelian
 multiplets under  both the SM gauge group and the hidden sector non-Abelian
 gauge group, thus preventing a clear distinction between the
 observable and the hidden sector of theory.

 \item
{\bf Effective Lagrangian.} In these models the couplings
 of the effective Lagrangian are calculable. In
particular, the conformal field theory techniques to  calculate
the Yukawa  couplings  as well as higher order couplings in the superpotential
  have been developed. 
However, in some cases 
detailed phenomenological analysis requires determination of 
  terms  of very high order (sometimes to
all orders) in the superpotential, 
 which is often beyond  calculable reach~\footnote{In general, one cannot 
classify the non-zero couplings just by using the gauge symmetries of the 
effective field theory, because certain gauge-allowed  terms are absent
due to the 
string dynamics.  In certain instances such constraints can be 
obtained by applying selection rules for the  corresponding string
amplitudes, as 
developed for orbifold \cite{CFTO} and blown-up orbifold compactifications
\cite{C}.  See also \cite{orbifolds}.}.
\end{itemize}

To address the phenomenology of this class of models,  the strategy is 
as follows~\footnote{Certain low energy phenomenological  consequences 
can be obtained by addressing only parts of the  analysis  described below. 
 In particular, taking this limited route,
 consequences for additional neutral gauge
bosons in this class of models were  extensively addressed in 
\cite{CL,CDEEL}.}:
\begin{itemize} 
\item
{\bf Step I.} Due to the anomalous $U(1)$ gauge symmetry 
at  genus-one there 
is an additional contribution  
 $\propto M_{string}^2/(192\pi^2){\rm Tr}Q_A$~\cite{DSW,ADS}
 to the corresponding $D$-term (Fayet-Iliopoulos term). 
 The contribution of 
  such a term is canceled \cite{DSW,ADS,DKI} by 
 giving non-zero  vacuum 
  expectation values (VEVs)   of ${\cal O}(M_{string})$ to certain 
massless  multiplets in such a way  that the anomalous $U(1)$ is broken, 
while
  maintaining the  $D$- and  $F$-flatness of  
  the  effective theory. Namely, a ``deficiency'' in the  string
  construction that produced the anomalous $U(1)$ is remedied at genus-one, 
   by providing  a  mechanism to  ``restabilize'' the vacuum. 
 A comprehensive analysis of the $D$- and $F$-flat directions for this class
 of models is  thus a prerequisite to address fully their phenomenological 
consequences.
 
 \item
 {\bf Step II.}  As the next step, the  surviving gauge group, the
massless 
particle
  quantum numbers and their couplings  in the effective Lagrangian at
$M_{string}$
  for the restabilized
 string vacua  should be analyzed.
 
 \item
{\bf Step III.} Finally, the implications of these models for the low
energy
phenomenology should be addressed. In this case one  faces the problem that 
the origin of the supersymmetry breaking 
 in string theory is not well understood,
and thus one is lacking predictive power for the supersymmetry breaking
parameters in the effective Lagrangian. Nevertheless,  one may parameterize 
 this  ignorance by 
  introducing  soft supersymmetry breaking  mass terms, and analyze the
 low energy consequences as a function of these  parameters. 
\end{itemize}

 The purpose of this paper is to describe a  systematic approach which 
 enables one to classify the  $D$- and $F$-flat directions which restabilize 
 the superstring vacuum (Step I).
  In particular, the analysis focuses on $D$- and $F$-flat
  directions with non-zero VEVs for the {\it non-Abelian singlets},
   only~\footnote{A 
number of
  specific examples of such $D$- and $F$-flat directions
   exist in the literature~\cite{orbifolds,fermionic}.  However, a
  comprehensive approach to this problem  is lacking.}. 
  (For the sake 
  of completeness an analysis of the same issue including non-Abelian 
  fields  as well
  is needed.
  However, we postpone this problem for further study.) 
While the method discussed in this paper is general, and applies to a general
perturbative heterotic string model with an anomalous $U(1)$,
 we chose to illustrate the  results for a particular
model, namely Model 5 of~\cite{CHL}. This  serves as an instructive 
 example, 
in which  the  classification of flat directions is
particularly simple.

In a 
subsequent paper~\cite{CCEEL}, 
the method is applied to the analysis of the flatness
 constraints 
 for a larger 
class of models.
In  later papers we plan to address the  phenomenological
consequences of these models in detail (Steps II and III).

This paper is organized as follows. In Section II some of the 
technical
features, in particular those associated with the anomalous $U(1)$ for
  the superstring models (based on the fermionic construction), are
discussed. In
Sections III  and  IV the approaches to determine $D$-flat directions for the 
{\it
non-anomalous}  $U(1)$'s  and  {\it anomalous} $U(1)_A$ are discussed, 
respectively.
 In Section V $F$-flatness constraints are analyzed. The summary and
conclusions are
given in Section VI.

\section{Stringy $SU(3)_C\times SU(2)_L\times U(1)_Y$ Models}

Gauge structures for quasi-realistic four-dimensional heterotic string
models follow the general form,
\beq
\{SU(3)_C\times SU(2)_L\in {\cal G} \}_{\rm obs}\times
\tilde{\cal G}^{\rm NA}_{\rm hid}\times 
\prod_n U(1)_n \times \prod_p \tilde{\Delta}_p\, , 
\label{ohs}
\eeq 
where ${\cal G}$ denotes a possible SM GUT or semi-GUT embedding
which may include some particular $U(1)$ factors, and  
$\tilde{\cal G}^{\rm NA}_{\rm hid}$ contains the non-Abelian hidden
sector gauge factors.
 Both observable and hidden $U(1)$ are included in
$\prod_n U(1)_n$, while $\prod_p \tilde{\Delta}_p$ 
denotes remaining local discrete symmetries.
  
``Cracks in the wall" separating the observable and hidden sectors
can appear in two forms. The first is when 
massless states appear as non-trivial representations of both the 
observable and hidden sector non-Abelian gauge groups (i.e, as ``mixed''
states). 
Model 5 of \cite{CHL} gives an example of such a mixed state that
carries
$SU(2)_{L}\times SU(2)_{\rm hid}$ charge. 
In this case part (or possibly all) of the so-called ``hidden sector''
is not actually hidden, unless the mixed states acquire string scale
masses.  If string scale masses are not acquired, a gauge-mediated process
may be the dominant form of observable sector supersymmetry breaking.

Second, there may be ``shadow'' sector $U(1)$'s, whose charges are carried
by both observable and hidden sector states. Although states with both 
shadow and hidden sector charge are generically present
in free fermionic models, acceptable  
phenomenology requires that such states have masses $m\ge m_{\rm susy}$.
Furthermore, traditional supergravity-mediated supersymmetry breaking 
suggests that such states should have masses $m\approx M_{\rm string}$.
Both types of communication between observable and hidden sectors can
have profound phenomenological implications.

The SM hypercharge, $Y$, may be totally embedded in a (semi-)GUT 
$\cal G$ or may also have
contributions from various $U(1)_n$. 
Typically, some of the extra $U(1)_n$ in four-dimensional string models
are anomalous. By this we mean those Abelian symmetries 
have mixed $U(1)$-gravitational anomalies and thus have a non-zero 
charge trace. 
For example, to date all known free fermionic four-dimensional 
$SU(3)_C\times SU(2)_L\times U(1)_Y$ models with extra, unembedded, 
$U(1)$ factors contain this type of  
anomaly.\footnote{Recently, though, an anomaly-free 
semi-GUT with 
$SU(4)_C\times SU(2)_L\times SU(2)\times U(1)^4\times
[ E_7 \times SU(2) ]_{\rm hid}$ gauge group containing  
observable sector chiral reps was found \cite{cf}. For a discussion on
anomalous $U(1)$'s in orbifold compactifications see \cite{kona}.}

In the initial charge basis, generally more than one $U(1)_n$ will appear 
anomalous. However, 
the total anomaly from all $U(1)_n$ can be rotated into a single 
$U(1)_{\rm A}$, defined by 
\beq
         U(1)_{\rm A} \equiv c_A\sum_n \{\Tr Q_{n}\}U(1)_n,
\label{rotau1}
\eeq
with $c_A$ a normalization coefficient.
Following rotation (\ref{rotau1}), an orthogonal basis $\{U(1)_a\}$
may be chosen  
for the non-anomalous components of the original set of $U(1)_n$.

As a result of the charge trace relationships,
generally known as the universal Green-Schwarz (GS) relation,
\beq
\frac{1}{k_mk_A^{1/2}}\mathop{\Tr}_{G_m}\, 
T(R)Q_A=\frac{1}{3k_A^{3/2}}\Tr Q_A^3
                        = \frac{1}{k_ak_A^{1/2}}\Tr Q_a^2Q_A 
                      =\frac{1}{24k_A^{1/2}}\Tr Q_A
                           \equiv 8\pi^2 \delta_{\rm GS} \, ,
\label{gs}
\eeq
invoked by stringy modular invariance constraints, 
rotation (\ref{rotau1}) removes all $U(1)$ triangle 
anomalies\footnote{This rotation is not necessarily sufficient to
remove all triangle anomalies from the $U(1)$ orthogonal to $U(1)_A$ 
for general field theoretic models or strongly coupled strings,
since the GS relations (\ref{gs}) need not hold then.}
 except those involving $U(1)_A$ 
\cite{DSW}.     
Here $k_m$ is the level of the gauge group $G_m$ and
$2T(R)$ is the index of the representation $R$. The physical content of
eq.~(\ref{gs}) is that the related mixed anomalies are canceled by the
pseudoscalar partner of the dilaton, which couples universally to all
gauge groups.  A similar set of anomaly-free $U(1)_a$ trace relations
indicates that triangle anomalies involving $Q_A^2$ cannot occur,
\beq
\frac{1}{k_mk_a^{1/2}}\mathop{\Tr}_{G_m}\, 
T(R)Q_a=\frac{1}{3k_a^{3/2}}\Tr Q_a^3
                     = \frac{1}{k_Ak_a^{1/2}}\Tr Q_A^2 Q_a 
                =\frac{1}{24k_a^{1/2}}\Tr Q_a = 0 \, .
\label{gsna}
\eeq
Relations analogous to $(\ref{gs})$ and $(\ref{gsna})$ involving
$Q_a Q_b$, $Q_a Q_b Q_c$, etc., also hold.

After rotation of all anomalies into a single $U(1)_A$ 
via (\ref{rotau1}), 
the standard anomaly cancelation mechanism
\cite{DSW} breaks $U(1)_{\rm A}$. However, this occurs 
at the expense of generating a Fayet-Iliopoulos (FI) $D$-term,
\beq
      {\e}^\phi \MP^2 \delta_{\rm GS}=
 \frac{{\e}^\phi \MP^2}{192\pi^2k_A^{1/2}}\Tr Q_A\, ,
\label{fid}
\eeq
where $\phi$ is the dilaton.
$ g\equiv {\e}^{\phi/2}$ 
is the physical four-dimensional gauge coupling.
The FI $D$-term is a genus-one string effect 
(one-loop effect in the effective field theory)
and is therefore calculable in perturbative string theory.
It must be canceled by 
appropriate VEVs of the scalar components $\varphi_i$ of supermultiplets
$\Phi_i$ carrying the anomalous charge so that\footnote{Our convention
for defining $D_{\rm A}$ is that the corresponding $D$ term in the
Lagrangian is $\frac{1}{2k_A}g^2D_{\rm A}^2$, and similarly for $D_a$.},
\beq
D_{\rm A} = \sum_i Q^{(A)}_i |\varphi_{i}|^2 
+ \frac{g^2 \MP^2}{192\pi^2}{\rm Tr}Q_{\rm A}
= 0\,\, .
\label{daf}
\eeq
$D$-flatness in the non-anomalous directions, along with $F$-flatness,  
impose additional constraints on the allowed directions of these VEVs,
\beqn
D_{a} &=& \sum_{i} Q^{(a)}_{i}|\varphi_{i}|^2 = 0,
\label{dif}\\
F_{i} &=& 
\frac{\partial W}{\partial \Phi_{i}} = 0;
\,\, W  =0.
\label{ff}
\eeqn

The presence of an anomalous $U(1)_A$ can have profound effects on a model 
including:
(i) Generation of a Fayet-Iliopoulos $D$-term and gauge rank reduction via
induced VEVs canceling the FI term \cite{orbifolds}. (ii) The induced VEVs can
eliminate many particles from the low energy spectrum.
(iii) It can play a role in explaining fermion mass hierarchies
\cite{anfer} and addressing the $\mu$ problem \cite{anmu}.  
Other possible implications are discussed in \cite{ancos,ansub,angam,answ2}. 

The model we use to illustrate our general analysis is model 5 of
ref.~\cite{CHL}, to which we refer for more details. The gauge group is
\be
\{SU(3)_C\times SU(2)_L\}_{\rm obs}\times\{SU(4)_2\times SU(2)_2\}_{\rm hid}
\times U(1)_A\times U(1)^6,
\ee
and the particle content includes, besides the MSSM multiplets, additional
chiral superfields:
\bear
&&6 (1,2,1,1) + (3,1,1,1) + (\bar{3},1,1,1) + \nonumber\\
&&4 (1,2,1,2) +\nonumber\\
&&2 (1,1,4,1) + 10 (1,1,\bar{4},1) + 8 (1,1,1,2) + \nonumber\\
&& 5 (1,1,4,2) + (1,1,\bar{4},2) + 8 (1,1,6,1) + 3 (1,1,1,3)+\nonumber\\
&& 42 (1,1,1,1)\;\;,
\eear
where $(a,b,c,d)$ indicates the representation under $(SU(3)_C,SU(2)_L,
SU(4)_2,SU(2)_2)$. In Table~I 
we list the 45 (including three generations of $e^+$ and $\nu^c$) 
non-Abelian singlets of the
model with their $U(1)$ charges. A prime on a
superfield $S_i$ indicates the existence of another superfield $S_i'$
with the same charges. Some fields $S_j$ have a mirror copy $\bar{S}_j$ 
with opposite $U(1)$ charges, and are indicated with a $\surd$.

Phenomenological considerations lead to the hypercharge definition
\cite{CHL}
\be
\label{hyp}
Y=\frac{1}{96}(-8Q_2-3Q_3-8Q_4-Q_5+Q_6),
\ee
[normalized to give $Y$(quark doublet)$=1/6$]. In following sections we
will examine those flat directions involving only non-Abelian singlets
with $Y=0$.
 
\section{$D$-Flat Directions for the Non-anomalous $U(1)$'s}

Any flat direction must have vanishing $D$-flat terms for
all the non-anomalous $U(1)$ factors of the gauge group. We consider first
the space of such flat directions, leaving aside for the moment the
questions of vanishing anomalous $D$-term and $F$-flatness. In the rest of
this section then, flat direction will refer to a direction with
zero $D$-terms for the non-anomalous $U(1)$'s.

A powerful and convenient way of analyzing the moduli space of flat
directions in a given model makes use of the correspondence between the
holomorphic gauge-invariant monomials built out of the chiral fields and
the $D$-flat directions \cite{flat0,flat1,flat2,flat3}. In the absence of
a
superpotential
[and of an anomalous $U(1)$, see next section], the set of all such
monomials describes the variety of classical supersymmetric vacua of the
model. When the superpotential does not vanish, the moduli space is
described by imposing $F$-flatness constraints on the holomorphic
monomials. 

We restrict our discussion to those flat directions exciting only 
chiral superfields
$\Phi_i$ (with scalar components $\varphi_i$) 
not charged under any of the non-Abelian gauge groups of a given model.
Those fields generically carry non-zero $U(1)_a$ charges, [$Q_i^{(a)}$
for the field $\Phi_i$], where we
reserve the index $a=A$ for the anomalous $U(1)$ and the rest, 
($a=1,...,m$) to the
non-anomalous $U(1)$'s (that is, we work in the rotated basis).
The space of all field directions with zero $D$-terms for $a=1,...,m$ is
described by the set of holomorphic $U(1)_a$ gauge-invariant monomials
in the $\Phi_i$'s.
Such monomials are not necessarily invariant under the anomalous $U(1)$
symmetry, so that they will generically carry a non-zero $U(1)_A$ charge. 

It may be useful at this point to spell out in more detail how this
correspondence works in this simple $U(1)$ case and show explicitly how
the flat directions are parameterized in terms of field VEVs. Let $N$ be
the number of
chiral non-Abelian singlets and let us normalize all $U(1)$ charges so
that
they
are integers. The $m$ $D$-flatness constraints are
\be
\label{dconst}
D_a=\sum_iQ_i^{(a)}|\varphi_i|^2=0,\;\;\;\;(a=1,...,m).
\ee
Consider now a generic holomorphic invariant monomial (HIM) of the form
\be
P\equiv\Pi'_i\Phi_i=\Pi_i\Phi_i^{n_i},
\ee
where the prime indicates that the index $i$ can take the same value several
times ($n_i\ge 0$).
The requirement of $U(1)_a$ gauge invariance of $P$ reads
\be
{\sum_i}'Q_i^{(a)}=\sum_in_iQ_i^{(a)}=0.
\ee
It is immediately obvious that the choice of VEVs
\be
|\varphi_i|^2=n_i|\psi|^2,
\ee
satisfies $D_a=0$ automatically for arbitrary $|\psi|$.
If the monomial $P$ is actually built as the product of two other HIM's
\be
P=N_1N_2=\left[\Pi_i\Phi_i^{n_i^{(1)}}\right]
\left[\Pi_j\Phi_j^{n_j^{(2)}}\right],
\ee
the VEVs can be chosen as
\be
|\varphi_i|^2=n_i^{(1)}|\psi_1|^2+n_i^{(2)}|\psi_2|^2,
\ee
with both $|\psi_1|$ and $|\psi_2|$ as arbitrary parameters (giving a
multidimensional flat direction).
It is clear then how HIM's correspond to $D$-flat directions. The converse
is also true. Given a solution to the $m$ eqs.~(\ref{dconst}), suppose no
subset of the fields with non-zero VEVs satisfy the same system of
equations (if there are, then we analyze each subset in turn).
This assumption implies that the field VEVs are related to each other by a
relation of the form $
|\varphi_j|^2=r_j|\psi|^2$ with $r_j\ge 0$ 
and
\be
\sum_jQ_j^{(a)}r_j=0.
\ee
Each $Q_j^{(a)}$ is an integer number by our choice of normalization.
Taking $r_1$ as an integer and having assumed that the solution to the
$D$-flat constraints cannot be split in sub-solutions, a theorem of
elementary algebra \cite{theo} tells that the $r_j$'s must be rational.
Rescaling them as integers, we can then construct
the holomorphic invariant monomial
\be
P=\Pi_j\Phi_j^{r_j}
\ee
associated with the particular solution of the $D$-flat constraints.

If the original solution $P$ could be decomposed in subsets of solutions,
then
the corresponding monomial would be a product of simpler invariant
monomials, associated as above with each separate subsolution. Flat
directions which cannot be decomposed in such a way would be
one-dimensional
(they will depend on a single arbitrary VEV $\psi$).
As a general rule, the (complex) dimension of a flat direction $P$ will be
equal to the number of different fields excited minus the number of $D$
constraints, as given by ${\mathrm Rank}[Q_i^{(a)}]|_P$, where
$Q_i^{(a)}$ denotes the full $N\times m$ charge matrix and $|_P$
means that only the rows corresponding to fields present in $P$ are
selected to compute the rank. This
rank is the number of $U(1)$'s which are broken along the flat
direction $P$. In other words, the initial number of degrees of freedom,
as given by the number of different fields in $P$, equals the dimension of
the flat direction plus the number of Goldstone bosons:
\be
\label{dimP}
[\#\; {\mathrm different\;\; fields\;\; in}\;\; P] =
{\mathrm dim}\; P + [\#\; {\mathrm broken}\; U(1)'s].
\ee

We next discuss the dimension of the space of $D$-flat directions. The
starting field space is an $N$-dimensional complex space. Any
one-dimensional
HIM (or equivalently, any one-dimensional solution of the $D$-flat
constraints)
\be
P=\Pi_i\Phi_i^{r_i},\;\; (r_i\ge 0)
\ee
can be associated with a vector in this $N$-dimensional field space
proportional to
\be
\label{vP}
v_P=(r_1,...,r_i,...,r_N),
\ee
 with a constant of proportionality  equal to the free VEV
$|\psi|^2$.
It is then clear that any flat direction can be expressed as a linear
combination of such elementary vectors\footnote{Not every linear
combination of these elementary vectors will correspond to a flat
direction though, because of the constraint $r_i\ge 0$, which renders
the problem non-trivial. In a linear combination of these vectors,
$v=\sum_\kappa a_\kappa v_{P_\kappa}$, negative coefficients $a_\kappa$
are allowed provided the vector $v$  also has $r_i\equiv\sum_\kappa
a_\kappa r_{i\kappa}\ge 0$ (see example at the end of
this section).} and one can construct a basis
of that space by finding a collection $\{v_\alpha\}$ of linearly independent
one-dimensional flat
directions. That basis, which has a finite number $d$ of
elements ($d\le N$), will be most useful for the description of the space
of flat directions.

In general, the dimension of that space is equal to the dimension $N$ of
the original
field space minus the number of independent  constraints imposed by the
$m$ $D_a=0$ conditions (these are real conditions but $m$ phases are
fixed by gauge fixing).
 This would give in principle $d=N-m$ for the dimension of the moduli
space of flat directions.
If not all $D_a=0$ conditions on the $\Phi_i$ are independent [the number of
independent conditions is the rank of the $N\times m$ charge matrix
$Q_i^{(a)}$] the 
dimension of the moduli space can be larger. In general,
\be
d\equiv {\mathrm dim} {\cal M}\leq N - {\mathrm Rank}\left[ Q_i^{(a)}\right].
\ee
In some cases the inequality is not saturated. This can only happen if
under some $U(1)$ (or linear combination of them) all the non-Abelian
singlet fields
have
$Q_i\ge 0$ or $Q_i\le 0$. There is no way of constructing an HIM that
includes
those fields with non-zero charges under that particular $U(1)$ so that
they will
never enter in any flat-direction\footnote{As ${\mathrm Tr} U(1)^{(a)}=0$
over all the fields in the model, this class of fields exists only when we
restrict our attention to a particular sector of the model. Such fields
can enter in flat directions that excite fields outside that sector.}.
Such fields may as well be removed from
the list of $N$ fields for the purpose of discussing the space of flat
directions. After such fields have been removed, one ends up with $N^*\le
N$ fields and
\be
d\equiv {\mathrm dim} {\cal M}= N^* - {\mathrm Rank}\left[
Q_i^{(a)}\right]^*,
\ee
and ${\mathrm Rank}\left[
Q_i^{(a)}\right]^*$ ($\le {\mathrm Rank}\left[
Q_i^{(a)}\right]$) is the rank of the $N^*\times m^*$ charge matrix of the
$N^*$
remaining fields [which have non-zero charges under $m^*\le m$ $U(1)$'s].

Once this basis ($\{M_\alpha\}$ HIM's with $\alpha=1,...,d$) has been
constructed, any flat direction $P$ can be expressed in terms of the
$M_\alpha$'s as\footnote{$P^n$ by itself is exactly the same flat
direction as $P$; the power is only important when relating $P$ to other
HIM's.} 
\be
\label{genP}
P^n=\Pi_\alpha M_\alpha^{n_\alpha}
\ee
where  $n$ and $n_\alpha$ are integers, with $n\geq 1$ but $n_\alpha$ can
take both signs\footnote{An equivalent description of eq.~(\ref{genP}) is
that the vector $v_P$ associated with the direction
$P$ can always be expressed as a linear combination of the basis vectors
$v_\alpha$.}. Such a basis contains in a compact form all the
information required about the classical moduli space.

We will present now such a basis for our model.
In principle, this basis
is not uniquely determined and different choices can be made which are
more or less convenient for different purposes. One
simple procedure to generate the basis is to make an automated
search of HIM's, retaining only those which are independent, until the
number of them saturates the dimension of the moduli space. An alternative
and more systematic way of generating the basis is to form first
a basis of invariants for the first $U(1)$ group. These invariants are
then combined to form an invariant basis for the second $U(1)$ factor
and so on. This procedure is simple because the construction of invariants
of a single $U(1)$ is trivial. However, the elements of the final basis
generated in this way are usually more complicated. In table~II, we report
two different bases for our example model (for flat directions built out
of $Y=0$ singlet fields only). An element like $\langle 8, 1^2\rangle$
stands for the HIM 
\be
M_6=\Phi_8\Phi_1^2.
\ee
Basis A is constructed in such a way as to minimize the number (and power)
of the fields entering the basis elements. Basis B will be suitable
for some future discussions.
Of course, all the results concerning flat directions are independent of
which particular basis is used.

In our model, after selecting $Y=0$ singlet fields only, we have 20 fields
plus 10 copies, and ${\mathrm Rank}\left[
Q_i^{(a)}\right]=6-1$ (6 initial non-anomalous $U(1)$'s minus hypercharge,
as the hypercharge $D$-term vanishes trivially for $Y=0$ fields). 
The dimension of the $D$-flat space
is then $d= 25$. In the table, we show only those elements involving the
20 different fields, so that we have only 15 elements. The 10 remaining
elements not shown are obtained trivially by appropriate substitution of
some fields by
their copies.

To clarify the meaning of eq. (\ref{genP}), we give an example of a flat
direction which is not contained in the bases of Table~II and requires
some negative $n_\alpha$:
\be
P\equiv\langle 22,8,\bar{5}\rangle =\frac{M_{13}M_9M_4}{M_{11}M_2}
=\frac{\langle 15,8,\bar{3}\rangle \langle 22,18,3 \rangle \langle
5,\bar{5}\rangle}{\langle 18,15,5\rangle\langle 3,\bar{3}\rangle},
\ee
or, in terms of the vectors of the form (\ref{vP}):
\be
v_P=v_{13}+v_9+v_4-v_{11}-v_2.
\ee

Although knowledge of the basis of $D$-flat directions is very
convenient,
it proves useful to have, in addition, a list of all one-dimensional flat
directions. The number of them can be much larger than the dimension
$d$ of the moduli space but is finite. 
In particular, from (\ref{dimP}), it is clear that there is an upper limit
on the number of different fields that can appear in a one-dimensional
flat direction. Using such a superbasis, every flat direction can be
factorized in terms of one-dimensional elements in an expression similar
to (\ref{genP}) but with $n_\alpha\ge 0$. This simplifies considerably
the analysis of the moduli space as only trivial multiplications of
one-dimensional HIM's are involved and positivity constraints on
the VEV-squares are automatically satisfied. For example, one is usually
interested
in finding if particular flat directions exist with non-zero VEVs for some
given fields while keeping zero VEVs for some other fields.
In principle, the use of the $d$-dimensional basis is sufficient to answer
such questions, but it can become cumbersome in practice (especially for
large dimensional bases, as is usually the case). In contrast, by
simple inspection of the superbasis one is able to determine whether such
directions do exist. The usefulness of such a list will become
clear in the next sections, especially for the discussion of $F$-flatness.
In Table~III
we present such a superbasis for our model example. Only the 73
directions involving non-primed fields are shown as the rest are obtained
trivially from these. 

\section{Anomalous $U(1)$}

In the presence of an anomalous $U(1)$ one should 
further require the vanishing of
the corresponding $D$-term, which now includes a Fayet-Iliopoulos term:
\be
D_A(P)=\sum_iQ_i^{A}|\varphi_i|^2+\xi,
\ee
with
\be
\xi=\frac{g^2 \MP^2}{192\pi^2}{\rm Tr}\, Q_A\neq 0.
\label{xidef}
\ee

 For the analysis of the flat directions in the presence of an anomalous
$U(1)$ it
is useful to find subsets of chiral fields for which the anomalous $U(1)$ is
a linear combination of the non-anomalous $U(1)$'s, say
\be
Q_j^{A}=\overline{Q}_j\equiv\sum_{a=1}^{m}\alpha_aQ^{(a)}_j,
\ee
where the $\alpha_a$ can be chosen for convenience.
Obviously, such a relation cannot hold for all the chiral fields in the model.
We will choose the $\alpha_a$ trying to maximize the number of fields for
which $\overline{Q}$ equals the anomalous
charge $Q_A$. It is remarkable that in all the models we have
studied, a definition exists which
matches the anomalous charges of a very large number of fields. This
observation helps considerably in the analysis and search of flat
directions.

Defining the quantities
\be
\hat Q_j=Q_j^{A}-\overline{Q}_j,
\ee
we can classify all the chiral fields in three
different types, depending on the sign of $\hat Q_j$ :
\bear
&&\Phi^+_j,\;\; {\mathrm if}\;\; \hat Q_j>0,\nonumber\\
&&\Phi^0_j,\;\; {\mathrm if}\;\; \hat Q_j=0,\nonumber\\
&&\Phi^-_j,\;\; {\mathrm if}\;\; \hat Q_j<0.
\eear
We can extend the definition of $\hat{Q}$ from the fields to the flat
directions. If the flat direction $P$ is defined by (\ref{genP}), then
\be
\hat{Q}(P^n)=n\hat{Q}(P)=\sum_\alpha n_\alpha
\hat{Q}(M_\alpha)
\ee
and
\be
\hat{Q}(M_\alpha=\Pi_i\Phi_i^{n_i})=\sum_in_i\hat{Q}_i.
\ee
By  invariance of $P$ under the non-anomalous $U(1)$'s, $\hat{Q}(P)$
turns out to be precisely
the anomalous charge of $P$.

The sign of the Fayet-Iliopoulos term $\xi$ determines which fields can
actually
form a flat direction. The statement is: 

{\em If $\xi>0$ $(<0)$,
any flat direction must contain at least one of the fields $\Phi^-_j$
$(\Phi^+_j)$}.

The proof goes as follows: First, if the flat direction is of the form
\be
P^0=\Pi'_i\Phi^0_i,
\ee
one has
\be
D_A(P^0)=\sum_iQ_i^{A}|\varphi_i^0|^2+\xi
=\sum_i\sum_{a=1}^{m}\alpha_aQ^{(a)}_i|\varphi_i^0|^2+\xi
=\sum_{a=1}^{m}\alpha_aD_a(P^0)+\xi=\xi\neq 0.
\ee

For a flat direction of the form
\be
P^\pm=(\Pi'_i\Phi^0_i)(\Pi'_j\Phi^\pm_j),
\ee
one has instead
\be
D_A(P^\pm)=\sum_i\sum_{a=1}^{m}\alpha_aQ^{(a)}_i|\varphi_i^0|^2+\xi+
\sum_jQ_j^{A}|\varphi_j^\pm|^2=
\xi+\sum_j\hat Q_j|\varphi_j^\pm|^2.
\ee
If $\xi>0$, the only possible way of 
canceling the $\xi$ term is by having negative $\hat Q_j$'s. A similar
 consideration applies for $\xi<0$, which requires some positive
$\hat Q_j$.
To cancel the Fayet-Iliopoulos term, some free VEV in the flat direction
$P$ (corresponding to some field with the right sign of $\hat{Q}$)
is fixed in terms of $\xi$ and the dimensionality of $P$ drops by one. The
final dimension of $P$ would satisfy eq.~(\ref{dimP}) if the anomalous
$U(1)$ is counted among the broken $U(1)$'s.

If in some model all the fields which are singlets under the
non-Abelian group factors have the wrong value of the $\hat Q_j$ charges
to
form a flat direction out of them, the
 cancelation of the Fayet-Iliopoulos term is necessarily accompanied by
the spontaneous
breaking of
some of the non-Abelian gauge group factors: some non-zero VEV for fields 
charged under the non-Abelian groups are required.

In table I we also show the values of $\hat{Q}_j$ for our example model,
with $\overline{Q}$ defined as
\be
\overline{Q}_j=-\frac{1}{3}Q_j^{(6)}.
\ee
We see that the only fields with non-zero $\hat{Q}$ are $S_1$
($\hat{Q}=32$), $\overline{S}_1$ ($\hat{Q}=-32$) and
${S}_{19}^({}'{}^)$ ($\hat{Q}=-32$). Then, any flat
direction involving only $Y=0$ singlets must necessarily have a non-zero
$S_1$ VEV [in this model the trace of the anomalous $U(1)$
is negative].

In Table II we list also the $\hat{Q}$ values for the basis elements.
If all the $\hat{Q}$'s were zero, $Q^A$ would be a linear combination
of the other $U(1)$'s in the sector considered. If all $\hat{Q}$'s for 
the elements of the basis are either zero or have the wrong sign to
cancel $\xi$, one cannot conclude that no flat direction
exists with the appropriate sign of $\hat{Q}$: starting with a $\hat{Q}=0$
direction that contains two fields with opposite $\hat{Q}$, the bad-sign
field could be divided out using the wrong sign $\hat{Q}$ basis element.
Either by manipulating the elements of the basis or by direct inspection
of the full list of one-dimensional flat directions it is easy to find
all
possible one-dimensional (before canceling $\xi$) directions with the
correct $\hat{Q}$ to cancel the
Fayet-Iliopoulos term. In our example model, from Table~III one finds
a total of five such directions:
\bear
\label{pis}
P_1&\equiv &R_6=\langle 8,1^2\rangle\nonumber,\\
P_2&\equiv &R_{22}=\langle 14,\bar{6},3,1^2\rangle\nonumber,\\
P_3&\equiv &R_{23}=\langle 18,5,3,1^2\rangle\nonumber,\\
P_4&\equiv &R_{42}=\langle 14,5,\bar{4},3,1^2\rangle\nonumber,\\
P_5&\equiv &R_{43}=\langle 18,\bar{6},4,3,1^2\rangle.
\eear
To these, one should add those similar monomials obtained by replacing
some field by its copy (
$\overline{S}_{6}\rightarrow \overline{S}_{6}'$, $S_{14}\rightarrow
S_{14}'$,
$S_{18}\rightarrow
S_{18}'$,
$S_5\rightarrow S_5'$, $S_8\rightarrow S_8'$).
In general models, the number of
different $P_\alpha$'s can be large and even exceed the dimension of the
moduli space. To describe the set of all $P_\alpha$'s
it is enough to find a subset of independent ones. Combinations of the
$P_\alpha$'s in that subset
will generate all the $P_\alpha$'s. A convenient basis for all the
$D$-flat
directions  can be arranged that contains this basis for the $P_\alpha$'s
as a sub-basis and is completed by other, independent, elements with
zero or the
wrong sign of $\hat{Q}$. Such a basis for our model is presented as
basis~B in Table~II. 

The defining properties of the $P_\alpha$ directions are as follows:
first, they are one-dimensional (and thus cannot be factorized in simpler
HIM's), and second, they have $\hat{Q}$ of the correct sign. These two
properties
determine the following important result:

Every $D$-flat direction can be written in the form
\be
P^n=P_\alpha N
\ee
with $N$ some HIM (not necessarily of $\hat{Q}>0$)
and $P_\alpha$ one of the five special HIM's listed in (\ref{pis})
(or some version of them involving copies of the fields).

\section{$F$-flatness}

When the superpotential $W$ is non-vanishing, the effective potential
will
receive $F$-term contributions that can lift some of  the $D$-flat
directions
we have discussed so far. 
The conditions to maintain zero potential along a given $D$-flat direction
are 
\be
F_i\equiv\partial W/\partial \Phi_i=0\,;\;\; W=0\,,
\ee
for all $\Phi_i$ in the theory. 

Consider a $D$-flat direction
\be
P=\Pi_i'\Phi_i,
\ee
[with $\hat{Q}(P)\neq 0$ of sign opposite to that of $Tr Q_{A}$].
 There are two types of superpotential terms
that
can  lift this generic flat direction. First, there can be terms of the
form
\be
\label{wa}
W_A\sim\left(\Pi_{i\in P}'\Phi_i\right)^n,
\ee
where $\Pi_{i\in P}'\Phi_i$ is a gauge invariant holomorphic monomial
formed by a subset of the chiral fields that enter in $P$. These terms
will generate a non-zero potential along the flat direction $P$
(barring a
cancelation of different contributions to $\partial W_B/\partial\Phi_i$).
A flat direction will be called type-A if such terms are allowed by the
gauge symmetries.
Note that $P$ itself is not gauge invariant under the anomalous $U(1)$,
and so terms $W_A\sim P^n$ are forbidden by gauge symmetry.
This is in contrast with the situation in the absence of an anomalous
$U(1)$, in which case the HIM's associated with any flat direction are
truly gauge invariant and thus all flat directions are type-A.

Other terms that can lift the generic flat direction $P$ are of the form
\be
\label{wb}
W_B\sim\Psi\left(\Pi_{i\in P}'\Phi_i\right),
\ee
with  $\Psi\notin P$. This term would contribute to the potential the
dangerous piece $|\partial W_B/\partial\Psi|^2$ which would 
lift $P$ (again barring cancelations). This exhausts all possible
superpotential terms giving a non-zero
potential along $P$.
As we are examining in this paper the flat directions formed out of
non-Abelian singlets only, we can restrict our attention to the
superpotential for these fields (gauge invariance requires that $\Psi$ in
$W_B$ is also a non-Abelian singlet).
We will say a flat direction is type-B if only terms of the type $W_B$
are allowed by the gauge symmetries.
The condition $W=0$ could be violated only by $W_A$ terms.

Ideally, one would like to find all directions which are $F$-flat to all
orders of the nonrenormalizable superpotential. If the $D$-flat direction
is type-A, that would require that some non-gauge symmetry (e.g., some
string selection rule derived using conformal field theory) conspires to
forbid the infinite number of terms
of  $W_A$ type.
If, on the other hand, the flat direction is type-B, only a finite number
of $W_B$ terms can exist, and knowledge of the superpotential up
to some finite order in the nonrenormalizable terms is all that is 
required to 
prove $F$-flatness to all orders.

We will restrict our analysis to type-B flat directions (in doing so, we
may of course leave out some true flat directions, but proving so is a
difficult task). A given $D$-flat direction $P$ will be type-B if no
gauge invariant [including $U(1)_A$] holomorphic monomial can be built out
of the fields in $P$. Equivalently, for a type-B $D$-flat direction $P$,
one cannot write
\be
\label{badfact}
P^n=NN' \;\;\; (n\ge 1),
\ee
with $N$ a fully invariant ($\hat{Q}=0$) holomorphic monomial.

The classification of $D$-flat directions we made in the previous section
will prove most useful for finding all type-B flat directions, as we will
show now in the case of our particular model.
Recall that any $D$-flat direction can be written as
\be
\label{gen1}
P^n=P_\alpha{}^({}'{}^)N,
\ee
where the $P_\alpha$'s are listed in (\ref{pis}).
There are several cases to consider depending on the anomalous charge of
$N$.
\begin{itemize}
\item $N=1$. This gives just the list of $P_\alpha$ $D$-flat
directions. These are type-B directions\footnote{One way to
check this
is to compute ${\mathrm Rank}[Q_i^A,Q_i^{(a)}]|_{P_\alpha}$. 
If this rank equals the number of different fields in
$P_\alpha$ no 
gauge-invariant  monomial can be built out of the fields in
$P_\alpha$.
However, the existence of an invariant is not guaranteed if the rank is
smaller.} because they are one-dimensional, and so cannot
be factorized as in (\ref{badfact}).
\item  $N=\Pi_i'\Phi_i$ has $\hat{Q}=0$. Then, $N$ is  gauge
invariant under all $U(1)$'s,  and the superpotential may contain
terms of the form
\be
W_A\sim N^n.
\ee
Thus, $P$ would be type-A.
\item  $N=\Pi_i'\Phi_i$ has $\hat{Q}<0$. To have $\hat{Q}(P)>0$ in
(\ref{gen1})
it is necessary that $|\hat{Q}(N)|<$$\hat{Q}(P_\alpha)$. As both
$\hat{Q}$'s
can be taken to be integers, the superpotential could contain terms of the
form
\be
W_A\sim\left(P_\alpha^mN^p\right)^n
\ee
with $m\hat{Q}(P_\alpha)+p\hat{Q}(N)=0$. This proves $P$ is type-A.
\item  $N=\Pi_i'\Phi_i$ has $\hat{Q}>0$. Then, using the results of
the previous section, we can write again
\be
N=P_\beta N'
\ee
and we should next analyze $N'$.
\end{itemize}

By getting back to the starting point in this way we have proved that all
type-B $D$-flat directions must be generated by combining 
the $P_\alpha$'s alone. 

We turn then to the analysis of all possible flat directions built out of
the $P_\alpha$'s in our model. Some combinations of $P_\alpha$'s will
not be type-B, and we will
not consider them further. For the type-B combinations it is simple to
find all possible $W_B$ type terms and the superbasis is very useful for
this purpose. For a given type-B direction $P$, any possible $W_B$ term
is built of the few elements in the superbasis which involve at most
one field not contained in $P$. Usually, it is enough to consider
one-dimensional $\hat{Q}=0$ invariants, as multidimensional invariants
will generally contain at least two fields (or the same field $\Psi$ to a
power larger than one) that do not appear in the type-B flat direction,
and thus do not spoil $F$-flatness.
By checking whether these terms
appear
in the superpotential (or are forbidden by stringy arguments) we can
determine whether the type-B $D$-flat direction is also $F$-flat to all
orders.
We then need to know the superpotential to some given order. In
ref.\cite{CHL}, $W$ was calculated up to $4^{th}$ order terms. The
non-Abelian singlet part reads\footnote{Those terms in
Table~III which have $\hat{Q}=0$ but do not appear in $W$ are examples of
terms forbidden
by conformal selection rules.}:
\bear
W_2&=&0,\vspace{0.1cm}\label{w2}\\
W_3&=&S_4(\overline{S}_{6}\overline{S}_{5}'
+\overline{S}_{6}'\overline{S}_{5})+
\overline{S}_{4}(S_{5}S_{6}'+S_{5}'S_{6})
+S_{18}(S_{5}S_{15}'+S_{5}'S_{15})\nonumber\\
&+&S_{14}(S_{15}\overline{S}_{6}'+S_{15}'\overline{S}_{6})
+S_{23}(S_{6}S_{8}'+S_{6}'S_{8})+
\overline{S}_{3}(S_{15}S_{8}'+S_{15}'S_{8})\label{w3}\\
&+&S_{22}(\overline{S}_{5}S_{8}'+\overline{S}_{5}'S_{8}),
\nonumber\vspace{0.1cm}\\
W_4&=&S_1S_8'S_{20}S_{19}'\label{w4}.
\eear
The different Yukawa couplings, of order $g$, are not indicated
explicitly, nor is the ${\cal O}(\MP^{-1})$ coefficient of $W_4$.
With this information we now examine all possible combinations
of the $P_\alpha$'s:
\begin{itemize}
\item $P_1=\langle 8,1^2\rangle$ and primed versions of it. This is type-B, 
and the only $W_B$ term allowed by gauge symmetries is the mass term
\be
W_B\sim S_1\overline{S}_1.
\ee
Mass terms do not appear in (\ref{w2}), and we conclude that $P_1$ is not only
$D$-flat but also $F$-flat to all orders in the nonrenormalizable terms in
the superpotential (the same applies to $P_1'=\langle 8',1^2\rangle$
and products of both).

\item $P_2=\langle 14,\bar{6},3,1^2\rangle$, etc. are also type-B,
with lifting terms
\bear
W_B^{(2)}&\sim 
&S_1\overline{S}_1+S_3\overline{S}_3+S_6\overline{S}_6+{\mathrm 
primed\; copies}\nonumber\\
W_B^{(3)}&\sim &S_{15}S_{14}\overline{S}_{6}+S_{23}S_{14}S_3 + {\mathrm
p.c.}
\eear
We see that $S_{15}'S_{14}\overline{S}_6$ does appear in (\ref{w3})
and lifts $P_2$. However, the directions $P_2'{}^({}'{}^)=\langle
14',\bar{6}{}^({}'{}^),3,1^2\rangle$ remain
flat to all orders (but $P_2P_2'$ and $P_2'''=\langle
14,\bar{6}',3,1^2\rangle$ are also lifted).

\item $P_3=\langle 18,5,3,1^2\rangle$ is type-B with (omitting mass terms)
\be
W_B^{(3)}\sim S_{18}S_{15}S_5+S_{22}S_{18}S_3 +{\mathrm p.c.}
\ee
Comparing with (\ref{w3}), we find that $S_{18}S_{15}'S_5$ lifts $P_3$
but $P_3'{}^({}'{}^)=\langle 18',5{}^({}'{}^),3,1^2\rangle$ are flat to
all orders (while $P_3P_3'$ and $P_3'''=\langle 18,5',3,1^2\rangle$
are also lifted).

\item $P_4=\langle 14,\bar{4},5,3,1^2\rangle$ and all its primed versions 
are type-B but are lifted by
\be
W\sim\overline{S}_4(S_5S_6'+S_5'S_6).
\ee

\item $P_5=\langle 18,\bar{6},4,3,1^2\rangle$ and all its primed versions 
are type-B but are lifted by
\be
W\sim S_4(\overline{S}_6\overline{S}_5'+\overline{S}_6'\overline{S}_5).
\ee

\item $P_1P_2=\langle 14,8,\bar{6},3,1^4\rangle$ is type-B. Besides those
terms already present for $P_1$ and $P_2$ separately, there are no
additional contributions to $W_B$. Then, only the combinations with
$P_2'{}^({}'{}^)$ remain
flat to all orders.

\item $P_1P_3$: same as above.

\item $P_2P_3=\langle 18,14,\bar{6},5,3^2,1^4\rangle$ is a 
particularly interesting type-B flat direction. The dangerous $W_B$ terms
are just those
discussed for $P_2$ and $P_3$. One way to satisfy the $F$-flatness
constraint is to choose adequately the primed copies appearing in the flat
direction. For example, $P_2'P_3'=\langle
18',14',\bar{6},5,3^2,1^4\rangle$ remains flat to all orders (because
$P_2'$ and $P_3'$ have zero
$F$-terms and there are no mixed terms to worry about).
Another interesting possibility is that there is a non-trivial
cancelation among different $F$-terms leaving some direction flat.
This happens for $P_2P_3=\langle 18,14,\bar{6},5,3^2,1^4\rangle$.
The term in the potential (with $\Psi=S_{15}'$) that would lift this flat
direction is
\be
\label{vf}
V_F=\left| \varphi_{14}\varphi_{\overline{6}}+\varphi_{18}\varphi_5\right|^2.
\ee
The $D$-term constraints would give the relations
\be
\begin{array}{lcl}
|\varphi_1|^2=2x^2,&\;\;\;\;& |\varphi_{\bar{6}}|^2=|\psi|^2,\nonumber\\
|\varphi_3|^2=x^2 ,&\;\;\;\;& |\varphi_{14}|^2=|\psi|^2,\nonumber\\
|\varphi_5|^2=x^2-|\psi|^2, &\;\;\;\;& |\varphi_{18}|^2=x^2-|\psi|^2,
\end{array}
\ee
with
\be
x^2=-\frac{\xi}{64},
\ee
and $|\psi|^2$ arbitrary (except for the constraint $|\psi|^2<x^2$).
It is trivial to see that with a convenient choice
of the sign of the VEVs in (\ref{vf}) and fixing
$|\psi|^2=x^2/2$, one obtains $V_F=0$ and thus $F$-flatness to all orders.
The same mechanism leaves $P_2'''P_3'''$ flat but cannot work for
$P_2P_3'''$ or $P_2'''P_3$.

\item $P_1P_2P_3=\langle 18,14,8,\bar{6},5,3^2,1^6\rangle$ is type-B.
The $W_B$ terms are those of $P_1$, $P_2$ and $P_3$. One possibility 
to obtain $F$-flatness is to choose conveniently  the primed copies and,
for example, $P_1P_2'P_3'=\langle 18',14',8,\bar{6},5,3^2,1^6\rangle$ 
remains $F$-flat.

As in the previous example, there is also the possibility of a cancelation
among different $F$-terms. In general, $P_1P_2P_3$ would be lifted
by the
same potential terms in
(\ref{vf}), but some particular choice of VEVs leaves $V_F=0$. More
precisely, solving the $D$-term
constraints gives
\be
\begin{array}{lcl}
|\varphi_1|^2=2x^2, &\;\;\;\;& |\varphi_8|^2=x^2-|\psi_1|^2,\nonumber\\
|\varphi_3|^2=|\psi_1|^2, &\;\;\;\; & |\varphi_{14}|^2=|\psi_2|^2,\nonumber\\
|\varphi_5|^2=|\psi_1|^2-|\psi_2|^2, &\;\;\;\; &
|\varphi_{18}|^2=|\psi_1|^2-|\psi_2|^2.\nonumber\\
|\varphi_{\bar{6}}|^2=|\psi_2|^2,&\;\;\;\;&
\end{array}
\ee
The choice $|\psi_1|^2=2|\psi_2|^2$ will give $V_F=0$ when the VEVs have
the proper signs.
In this case, we are left with a one-dimensional  direction
(parameterized
by, say, $|\psi_2|^2$)
$F$-flat to all orders. The particular point $|\psi_2|^2=x^2/2$ gives
$|\varphi_8|^2=0$ and corresponds
to the flat direction $P_2P_3$ discussed before.

\item $P_4(\Pi_i P_\alpha)$. The structure of the superpotential terms
that lift $P_4$ is such
that this type of $D$-flat directions is always lifted (a cancelation of
different $F$-terms as in
 the examples above cannot be enforced). Moreover, some of the possible
combinations, like $P_4P_5$,
are type-A as they contain the invariant $\langle 6,\bar{6}\rangle$.

\item $P_5 (\Pi_i P_\alpha)$. Same as above.
\end{itemize}

A list of type-B flat directions which remain flat to all orders is
presented in Table~IV. The second column gives the
 dimensionality (a zero entry means that all field VEVs are determined in
terms of the Fayet-Iliopoulos
 term) and the second gives the number of non-anomalous $U(1)$'s broken
along the
corresponding direction. Those
 directions which are flat after imposing $F$-term constraints
are indicated by $|_F$. All type-B flat directions are trivially
built out of the zero-dimensional ones in Table~IV by 
addition of the primed copies of some fields. Some examples are already
shown in that Table. Addition of a primed field to a given direction has
the effect of increasing its dimensionality by one while the
number of broken $U(1)$'s is the same. For example, the addition of $S_8'$
to $P_1$ results in the direction $P_1P_1'$ and, as is shown in Table IV, 
the above rule is verified.

We can summarize our results by noting that all of the above flat
directions are particular cases of the type-B flat direction
\be
\label{bigP}
P_1P_1'P_2'P_2'''P_3'P_3'''|_F.
\ee
This direction is five-dimensional and all other type-B flat directions
can be viewed as particular sub-spaces of it, generated by setting some of
the VEVs in (\ref{bigP}) to zero or imposing some extra relation among
them. 

In other words, the moduli space of type-B flat directions is a
subspace of the twelve-dimensional field space
\be
\{S_1,S_3,S_5,S_5',\overline{S}_6,\overline{S}_6',
S_8,S_8',S_{14},S_{14}',S_{18},S_{18}'\}
\ee
obtained by imposing the constraints
from $D$-flatness (4 independent non-anomalous constraints plus the
anomalous
one) and $F$-flatness (the 2 conditions 
$\varphi_{14}\varphi_{\overline{6}}+\varphi_{18}\varphi_{5}=0$ and
$\varphi_{14}\varphi_{\overline{6}}'+\varphi_{18}\varphi_{5}'=0$).

\section{Summary and Conclusions}

In this paper,  we have developed techniques to  classify systematically
 the  $D$- and $F$-flat directions  for  perturbative heterotic superstring 
vacua   with  $N=1$ supersymmetry and an
 anomalous $U(1)$ in the gauge group. At   genus-one  the  superstring
theory
generates a Fayet-Iliopoulos $D$-term 
 $\xi\propto M_{string}^2/(192\pi^2){\rm Tr}Q_A$~\cite{DSW,ADS}.
The presence of this term triggers non-zero
vacuum expectation values (VEVs) for certain
massless multiplets in such a way  that the anomalous $U(1)$ is broken
while maintaining the  $D$- and  $F$-flatness of the effective  theory.
Thus, the superstring vacuum is restabilized~\cite{DSW,ADS,DKI} and
supersymmetry remains unbroken.
  
 The analysis  developed in this paper focuses on classification of the
 $D$- and $F$-flat
  directions  involving the {\it non-Abelian singlets} only; the 
results are illustrated   with the  explicit construction of flat directions 
for a particular string 
model, i.e.,  Model 5 of~\cite{CHL}, which is
 based on the  free fermionic construction.

The analysis was  set up along the following stages:
\vskip 5mm
{\noindent I. \bf $D$-flatness}
\vskip 5mm
\begin{itemize}
\item
{\bf Non-anomalous $U(1)$'s.}
First, the $D$-flat directions associated with the non-anomalous $U(1)$'s are
determined, by making use of the one-to-one correspondence of the $D$-flat
directions with the  holomorphic
gauge-invariant monomials (HIM) of  chiral superfields, constructed from
the
non-Abelian singlets~\footnote{The fact that
 each chiral superfield enters the HIM's with a {\it positive power} is 
equivalent to
 ensuring that a particular $D$-flat direction 
 involves VEVs  of  the  fields whose {\it absolute value-squared is 
 positive}.}. We  determined the moduli
space of  all such flat directions. Its dimension is
$d\equiv {\rm dim}{\cal M}= N^*-{\rm Rank} \left[Q_i^{(a)}\right]^*$, 
where $N^*$
is the number of  SM singlet fields that enter  HIM's, and 
${\rm Rank} \left[Q_i^{(a)}\right]^*$ is the rank of the $N^*\times m^*$ 
charge matrix  of the  $N^*$ fields, which have non-zero charges under $m^*$
non-anomalous $U(1)^{(a)}$ factors. (The latter matrix is
the one associated with  $m^*$ $D_{(a)}$-flatness
conditions for $N^*$ fields).

The construction of all flat directions from a basis of $d$ independent HIM's 
that characterizes the moduli space requires both multiplication and
division of 
the elements of the basis, making it cumbersome to scan the whole moduli
space.  This 
difficulty is avoided if one considers another useful set, referred to as a  
superbasis,
 which is the set of {\it all 
one-dimensional}  HIM's (see Table III for  the chosen model).
 Every  $D$-flat direction can
then be obtained as a {\it product}  of elements in this superbasis, so that 
positivity of the VEV squares is automatic and the contents of the moduli
space are more clearly displayed.  Such a set can
be obtained  by an automated computer search.
\item
{\bf Anomalous $U(1)$.} The  flatness of the
 $D$-term associated with the anomalous $U(1)$
requires that the Fayet-Iliopoulos term  $\xi $ is balanced against the 
contribution to $D_A$ from the non-Abelian singlets $\varphi_i$, such that 
 $D_A= \sum_iQ^A_i|\varphi_i|^2+\xi=0$.
 We  classify the HIM's  according to  the sign of their contribution  
to the $D_A$-term, that is, according to their anomalous charge. They fall
into three classes: their anomalous charge has a sign  
(i) equal to $-{\rm sign}\; \xi$,  (ii) zero, (iii)  equal to $+{\rm
sign}\; \xi$.
 To ensure the $D_A$-flatness constraint,
the (super)basis should  necessarily  contain one or more HIM's  belonging
to class (i).
If this is not the case
 then  the $D_A$-flatness constraint 
 cannot be satisfied,  i.e.,  there is a ``no-go theorem'' for the
restabilization  of  the string vacuum
via  VEVs of  the  non-Abelian singlets\footnote{Whether or not this is 
the case can be discerned already at the level of the elementary fields, 
without the need of computing the superbasis, as is discussed in 
section~IV.}. In this case, the procedure
 necessarily requires non-zero VEVs for fields that transform under the
non-Abelian gauge factors, further reducing the rank of the gauge group.

 $D_A$-flat directions  must  therefore contain elements of the
 the superbasis  that belong  to class (i).
However, they  can be multiplied  by
 the elements of the basis  in class (ii) and/or
class(iii), as long as  the resulting HIM also belongs to class (i).

\end{itemize} 
 
 \vskip 5mm
{\noindent II. \bf F-flatness}
\vskip 5mm
$F$-flatness  requires  $\partial W/\partial\Phi_i=0$, $W=0$ for all
massless chiral 
superfields $\Phi_i$ of the model. 
 The  $D$-flat 
directions can be lifted  due to
the   following 
two types of terms in the superpotential:
\begin{itemize}
  \item
  {\bf W$_{\bf A}$-Superpotential.} Those are terms in the superpotential
that
  include only  the non-Abelian singlets which acquire non-zero VEVs
along some direction (while
  ensuring  the $D$-flatness of the  effective theory).  
  Products of elements of the superbasis which  belong to class (ii) (their
  contribution to  $D_A$ is zero) are {\it not  forbidden} by gauge invariance,
  and thus they may appear in $W_A$. 
  In certain cases one may be able to show that 
a stringy (world-sheet) symmetry  ensures that a particular HIM element (with
zero contribution to $D_A$)  and 
all its positive powers  are  absent. However, in general  conformal field
theory techniques may not be powerful enough  to determine the absence of
all such terms.  We therefore took a ``conservative'' approach 
that all  such  HIM elements could  appear in $W_A$ at some order  and 
 could thus lift the (type-A)
$D$-flat direction (barring cancelation among different terms of that
type).
Consequently, we remove the  elements in the superbasis that belong to class 
(ii).  (It is of course possible that there are additional flat directions that 
 are missed by our conservative approach). One can also show  (see Section
V) that 
 elements of the superbasis belonging to class (iii) 
should also be removed,  because they  can 
 generate  terms in class (ii) when they are multiplied  with
appropriate
powers of the superbasis elements in  class (i). 
Therefore, the remaining elements in the superbasis,  which  ensure
type-A  $F$-flatness to all orders,
 consist of {\it the subset belonging to class (i)} only.

\item
{\bf W$_{\bf B}$-Superpotential.} Those are the  terms in the
superpotential
 which contain {\it one power} of a
non-Abelian singlet field $\Psi_i$ with {\it zero} VEV, while  all the 
other fields have non-zero VEVs along the given direction.
Such terms, if  present,
 can lift (type-B) $F$-flat directions (i.e. those directions for which
no $W_A$ terms are possible). Gauge invariance constrains
 the number of  allowed  terms in $W_B$ to  a {\it finite}
 number.  By doing a  string calculation, one can then check  explicitly
 whether these terms are absent thus ensuring the type-B
$F$-flatness  conditions~\footnote{In certain cases the type-B $F$-flat
directions can also be obtained by  employing cancelation of different
  terms in $W_B$ (See Section V).}.
 
Combining class (i) elements of the superbasis and checking the
superpotential, one can generate all type-B directions which are $D$- and
$F$-flat to all orders.
 
\end{itemize}

In conclusion, we developed techniques
that  set the stage
 to  classify systematically
the $D$- and $F$-flat   directions involving nonabelian singlet fields
only for a 
large class of 
  perturbative string vacua
with an anomalous $U(1)$. 
In a subsequent paper~\cite{CCEEL}, these techniques are applied 
 to  a  class of models  based on the free
  fermionic construction,  which possess three ordinary families and 
 the standard model gauge
 group in the observable sector.
 
 \acknowledgments
This work was supported in part by U.S. Department of Energy Grant No. 
DOE-EY-76-02-3071. 
\newpage

\def\NPB#1#2#3{{\it Nucl.\ Phys.}\/ {\bf B#1} (19#2) #3}
\def\PLB#1#2#3{{\it Phys.\ Lett.}\/ {\bf B#1} (19#2) #3}
\def\PRD#1#2#3{{\it Phys.\ Rev.}\/ {\bf D#1} (19#2) #3}
\def\PRL#1#2#3{{\it Phys.\ Rev.\ Lett.}\/ {\bf #1} (19#2) #3}
\def\PRT#1#2#3{{\it Phys.\ Rep.}\/ {\bf#1} (19#2) #3}
\def\MODA#1#2#3{{\it Mod.\ Phys.\ Lett.}\/ {\bf A#1} (19#2) #3}
\def\IJMP#1#2#3{{\it Int.\ J.\ Mod.\ Phys.}\/ {\bf A#1} (19#2) #3}
\def\nuvc#1#2#3{{\it Nuovo Cimento}\/ {\bf #1A} (#2) #3}
\def\RPP#1#2#3{{\it Rept.\ Prog.\ Phys.}\/ {\bf #1} (19#2) #3}
\def\etal{{\it et al\/}}

\bibliographystyle{unsrt}

\newpage

\vskip .3truecm
\begin{center}
\begin{tabular}{|c|c|ccccccc|c|c|}
\hline\hline
NA     &Vector-   &$Q_A$&$Q_1$&$Q_2$&$Q_3$&$Q_4$&$Q_5$&$Q_6$
                  &$Q_{Y}$&$\hat{Q}$\\
Singlet&like      &&&&&&&&&\\ 
       &$+/-$     &&&&&&&&&\\ 
\hline\hline
$S_1$& $\surd$    &   28&   0&   0&   0&   0&  12&  12& 0  &32\\ 
$S_2$& $\surd$    &    8&   0&   4&  16&   0&  -8& -24&-1  &0\\
$S_3$& $\surd$    &    8&   0&   0& -16&   4&  -8& -24& 0  &0\\ 
$S_4$& $\surd$    &    0&   4&   0&   0&   0&   0&   0& 0  &0\\ 
$S_5{}^({}'{}^)$& $\surd^({}'{}^)$    &    0&   2&   4&   0&  -2& -16&   0& 0  &0\\ 
$S_6{}^({}'{}^)$& $\surd^({}'{}^)$    &    0&   2&  -4&   0&   2&  16&   0& 0  &0\\ 
$S_7$&            &   20&   0&  -2&   8&   2&  12& -60&-1  &0\\ 
$S_8{}^({}'{}^)$&            &    8&   0&   0&   0&   0& -24& -24& 0  &0\\ 
$S_{9 }$&         &    4&  -4&  -2& -24&  -2&  -4& -12& 1  &0\\ 
$S_{10}$&         &    4&   4&  -2& -24&  -2&  -4& -12& 1  &0\\ 
$S_{11}{}^({}'{}^)$&         &    4&   0&   2&   8&  -6&  -4& -12& 0  &0\\ 
$S_{12}$&         &    4&   0&   2& -24&   2&  28& -12& 0  &0\\ 
$S_{13}$&         &    0&   2&   0& -16&  -2& -32&   0& 1  &0\\ 
$S_{14}{}^({}'{}^)$&         &    0&   2&  -4&  16&  -2&   0&   0& 0  &0\\ 
$S_{15}{}^({}'{}^)$&         &    0&   0&   0& -16&   4&  16&   0& 0  &0\\ 
$S_{16}{}^({}'{}^)$&         &    0&   0&   4&  16&   0&  16&   0&-1  &0\\ 
$S_{17}$&         &    0&  -2&   0& -16&  -2& -32&   0& 1  &0\\ 
$S_{18}{}^({}'{}^)$&         &    0&  -2&  -4&  16&  -2&   0&   0& 0  &0\\
$S_{19}{}^({}'{}^)$&         &  -24&   0&   2&  -8&  -2&   0& -24& 0&-32\\ 
$S_{20}$&         &  -12&   0&  -2&   8&   2&  12&  36& 0  &0\\ 
$S_{21}{}^({}'{}^)$&         &   -8&   0&  -4&   0&  -4&  -8&  24& 1  &0\\ 
$S_{22}$&         &   -8&   2&   4&   0&  -2&   8&  24& 0  &0\\ 
$S_{23}$&         &   -8&  -2&   4&   0&  -2&   8&  24& 0  &0\\ 
$S_{24}{}^({}'{}^)$&         &   -4&   0&   2&  24&   2&   4&  12&-1  &0\\ 
$S_{25}{}^({}'{}^)$&         &   -4&   0&  -2&  -8&  -6&   4&  12& 1  &0\\ 
\hline
\hline
\end{tabular}
\end{center}
\noindent Table I: List of non-Abelian singlet fields in the model with
their charges under the $U(1)$ gauge groups, hypercharge as defined in
eq.(\ref{hyp}) and $\hat{Q}=Q_A+Q_6/3$. A prime in parentheses
indicates that there is an extra copy of the field with exactly the same
$U(1)$ charges. A $\surd$ in the second column indicates that there is an
extra copy of the field with exactly opposite $U(1)$ charges.

\newpage

\vskip .3truecm

\begin{center}
\begin{tabular}{|l|c||l|c|}
\hline\hline
BASIS A     &   $\hat{Q}$   & BASIS B & $\hat{Q}$ \\
\hline\hline
$M_1=\langle 1,\bar{1}\rangle$ &0 & $P_1=\langle 8,1^2 \rangle$ & 64 \\
$M_{2}=\langle 3,\bar{3}\rangle$&0&$P_2=\langle 14,\bar{6},3,1^2 \rangle$ &64\\
$M_{3}=\langle 4,\bar{4}\rangle$ &0&$P_3=\langle 18,5,3,1^2 \rangle$ & 64\\
$M_{4}=\langle 5,\bar{5}\rangle$&0&$P_4=\langle 14,5,\bar{4},3,1^2 \rangle$&64\\
$M_{5}=\langle 6,\bar{6}\rangle$ &0 
& $P_5=\langle 18,\bar{6},4,3,1^2 \rangle$ & 64\\
$M_{6}=\langle 8,1^2\rangle$    &64 
&$N_1=\langle 1,\bar{1}\rangle$ &0 \\
$M_{7}=\langle 20,11,3\rangle$ &0 
&$N_2=\langle 3,\bar{3}\rangle$ &0 \\
$M_{8}=\langle 23,14,3\rangle$ &0 
& $N_3=\langle 5,\bar{5}\rangle$ &0\\
$M_{9}=\langle 22,18,3\rangle$ &0 
& $N_4=\langle 6,\bar{6}\rangle$ &0\\
$M_{10}=\langle \bar{6},\bar{5},4\rangle$ &0
&$N_{5}=\langle 20,11,3\rangle$ &0 \\
$M_{11}=\langle 18,15,5\rangle$ &0 & $N_{6}=\langle 23,14,3\rangle$ &0\\
$M_{12}=\langle 23,8,6\rangle$ &0 & $N_{7}=\langle 22,18,3\rangle$ &0 \\
$M_{13}=\langle 15,\bar{3},8\rangle$ &0 & $N_8=\langle 18,15,5\rangle$ &0\\
$M_{14}=\langle 20,19,\bar{1}\rangle$ &-64
&$N_{9}=\langle 20,19,\bar{1}\rangle$ &-64\\
$M_{15}=\langle 20,18,12,8,5\rangle$ &0 
& $N_{10}=\langle 20,18,12,8,5\rangle$ &0 \\
\hline
\hline
\end{tabular}
\end{center}
\noindent Table II: Two different bases of the moduli space of
non-anomalous $D$-flat directions of the model. Each element corresponds
to
a holomorphic monomial which is gauge invariant under the non-anomalous
$U(1)$'s. The anomalous charge is given by $\hat{Q}$. 

\newpage

 \vskip .3truecm
\begin{center}
\begin{tabular}{|l|c||l|c||l|c|}
\hline\hline
SUPERBASIS      &   $\hat{Q}$   &  
$R_{25}=\langle 23,\bar{5},4,\bar{1}^2\rangle$ &-64 &
$R_{50}=\langle 22,14,\bar{6},\bar{5},3\rangle$ &0   \\
$R_1=\langle 1,\bar{1}\rangle$ &0 & 
$R_{26}=\langle 22,6,\bar{4},\bar{1}^2\rangle$ &-64& 
$R_{51}=\langle 23,18,6,5,3 \rangle$ & 0 \\
$R_{2}=\langle 3,\bar{3}\rangle$ &0 & 
$R_{27}=\langle 20^2,12,11,\bar{1}^4\rangle$ &-128 & 
$R_{52}=\langle 20,18^2,12,5^2,3\rangle$ & 0\\
$R_{3}=\langle 4,\bar{4}\rangle$ &0 & 
$R_{28}=\langle 20,15,11,\bar{1}^2\rangle$ &-64& 
$R_{53}=\langle 20^2,19^2,18,5,3 \rangle$ & -64\\
$R_{4}=\langle 5,\bar{5}\rangle$ &0 &  
$R_{29}=\langle 23,15,14,\bar{1}^2\rangle$ &-64 & 
$R_{54}=\langle 20,14^2,12,\bar{6}^2,3 \rangle$ & 0 \\
$R_{5}=\langle 6,\bar{6}\rangle$ &0 &  
$R_{30}=\langle 22,18,15,\bar{1}^2 \rangle$ &-64 &
$R_{55}=\langle 20^2,19^2,14,\bar{6},3 \rangle$ & -64\\
$R_{6}=\langle 8,1^2\rangle$    &64 &  
$R_{31}=\langle 22,14,\bar{4},3 \rangle$ &0&
$R_{56}=\langle 22,15,14,8,\bar{4}\rangle$ &0 \\
$R_{7}=\langle 15,\bar{3},\bar{1}^2\rangle$ &-64 &  
$R_{32}=\langle 23,18,4,3\rangle$ &0 & 
$R_{57}=\langle 23,18,15,8,4\rangle$ &0 \\
$R_{8}=\langle 22,\bar{5},\bar{1}^2\rangle$ &-64 &  
$R_{33}=\langle 15,14,5,\bar{4} \rangle$ &0  & 
$R_{58}=\langle 20,18,12,8,5\rangle$ &0\\
$R_{9}=\langle 23,\bar{6},\bar{1}^2\rangle$ &-64 &  
$R_{34}=\langle 18,15,\bar{6},4\rangle$ &0&
$R_{59}=\langle 20,14,12,8,\bar{6}\rangle$ &0\\
$R_{10}=\langle 20,19,\bar{1}\rangle$ &-64&
$R_{35}=\langle 20,12,8^2,\bar{3} \rangle$ & 0& 
$R_{60}=\langle 23,20,14,12,8^2\rangle$ &0 \\
$R_{11}=\langle 20,11,3\rangle$ &0 &  
$R_{36}=\langle 23,8,\bar{5},4\rangle$ & 0 & 
$R_{61}=\langle 22,20,18,12,8^2\rangle$ &0\\
$R_{12}=\langle 23,14,3\rangle$ &0 &  
$R_{37}=\langle 22,8,6,\bar{4}\rangle$ & 0 & 
$R_{62}=\langle 20,14,12,5,\bar{4},\bar{1}^2\rangle$ &-64 \\
$R_{13}=\langle 22,18,3\rangle$ &0 & 
$R_{38}=\langle 20^2,12,11,8^2 \rangle$& 0 &
$R_{63}=\langle 20,18,12,\bar{6},4,\bar{1}^2\rangle$ &-64\\
$R_{14}=\langle \bar{6},\bar{5},4\rangle$ & 0  &  
$R_{39}=\langle 20,15,11,8\rangle$ &0& 
$R_{64}=\langle 22,20,14,12,\bar{4},\bar{1}^4\rangle$ &-128\\
$R_{15}=\langle 6,5,\bar{4}\rangle$ &0 &  
$R_{40}=\langle 23,15,14,8\rangle$ &0&
$R_{65}=\langle 23,20,18,12,4,\bar{1}^4\rangle$ &-128 \\
$R_{16}=\langle 18,15,5\rangle$ &0&
$R_{41}=\langle 22,18,15,8\rangle$ &0&
$R_{66}=\langle 20,14^2,12,5^2,\bar{4}^2,3\rangle$ &0 \\
$R_{17}=\langle 15,14,\bar{6}\rangle$ &0&
$R_{42}=\langle 14,5,\bar{4},3,1^2\rangle$ &64&
$R_{67}=\langle 20^2,19^2,14,5,\bar{4},3\rangle$ &-64 \\
$R_{18}=\langle 15,8,\bar{3}\rangle$ &0&
$R_{43}=\langle 18,\bar{6},4,3,1^2\rangle $&64&
$R_{68}=\langle 20,18^2,12,\bar{6}^2,4^2,3\rangle$ &0 \\
$R_{19}=\langle 22,8,\bar{5}\rangle$ &0&
$R_{44}=\langle 22,15,14,\bar{4},\bar{1}^2\rangle$ &-64&
$R_{69}=\langle 20^2,19^2,18,\bar{6},4,3\rangle$ &-64 \\
$R_{20}=\langle 20^2,19^2,8\rangle$ &-64&
$R_{45}=\langle 23,18,15,4,\bar{1}^2\rangle$ &-64&
$R_{70}=\langle 20,14,12,8,5,\bar{4}\rangle$ &0 \\
$R_{21}=\langle 23,8,6\rangle$ &0&
$R_{46}=\langle 20,18,12,5,\bar{1}^2\rangle$ &-64&
$R_{71}=\langle 20,18,12,8,\bar{6},4\rangle$ &0 \\
$R_{23}=\langle 14,\bar{6},3,1^2\rangle$ &64&
$R_{47}=\langle 20,14,12,\bar{6},\bar{1}^2\rangle$ &-64&
$R_{72}=\langle 22,20,14,12,8^2,\bar{4}\rangle$ &0 \\
$R_{22}=\langle 18,5,3,1^2\rangle$ &64&
$R_{48}=\langle 23,20,14,12,\bar{1}^4\rangle$ &-128&
$R_{73}=\langle 23,20,18,12,8^2,4\rangle$ &0 \\
$R_{24}=\langle 20,12,\bar{3},\bar{1}^4\rangle$ &-128&
$R_{49}=\langle 22,20,18,12,\bar{1}^4\rangle$ &-128&
$\;\;\;\;\;$$--------$&- \\
\hline
\hline
\end{tabular}
\end{center}
\noindent Table III: Complete list of one-dimensional non-anomalous
$D$-flat
directions of the model with their corresponding anomalous charges. 

\vskip 1.truecm 
\def\stupidlatex{    \bar{6}{}^({}'{}^)   }

\begin{center}
\begin{tabular}{|l|c|c|}
\hline\hline
FLAT DIRECTION    &   Dim.  & $\#$ $U(1)$'s  \\
\hline\hline
$P_{1}{}^({}'{}^)=\langle 8{}^({}'{}^),1^2\rangle$    & 0 & 1 \\
$P_{2}'{}^({}'{}^)=$$\langle 14',\stupidlatex,3,1^2\rangle$ & 0 & 3 \\
$P_{3}'{}^({}'{}^)=\langle 18',5^({}'{}^),3,1^2\rangle$ & 0 & 3 \\
$P_2P_3|_F$, $P_2'''P_3'''\equiv \langle 18,14,
\bar{6}',5',3^2,1^4\rangle|_F$ & 0 & 4 \\
$P_{1}P_1'$    & 1 & 1 \\
$P_{2}'P_2''$, $P_{3}'P_3''$    & 1 & 3 \\
$P_1{}^({}'{}^)P_2'{}^({}'{}^)$, $P_1{}^({}'{}^)
P_3'{}^({}'{}^) $ & 1 & 3\\
$P_2'{}^({}'{}^)P_3'{}^({}'{}^)$ & 1 & 4 \\
$P_1{}^({}'{}^)P_2P_3|_F$, $P_1{}^({}'{}^)P_2'''P_3'''|_F$ & 1 & 4\\
\hline
\hline
\end{tabular}
\end{center}
\noindent Table IV: List of type-B $D$-flat directions which are
$F$-flat to all orders for the model discussed. The dimension of the
direction, after
cancelation of the Fayet-Iliopoulos term, is indicated in the second
column. The third column gives the number of non-anomalous $U(1)$'s
broken along the flat direction. 
\end{document}